\documentclass[manuscript,screen,nonacm]{acmart} 
\usepackage{natbib}

\AtBeginDocument{%
  \providecommand\BibTeX{{%
    \normalfont B\kern-0.5em{\scshape i\kern-0.25em b}\kern-0.8em\TeX}}}

\setcopyright{acmcopyright} %{none} removes copyright
\setcopyright{none}
\copyrightyear{2021}
\acmYear{2021}
\acmDOI{10.1145/1122445.1122456}

\acmConference[CHI '21 Extended Abstracts]{CHI EA '21: Extended Abstracts of the 2021 CHI Conference on Human Factors in Computing Systems}{May 08--13, 2021}{Yokohama, Japan}
\acmBooktitle{CHI EA '21: Extended Abstracts of the 2021 CHI Conference on Human Factors in Computing Systems,
  May 08--13, 2021, Yokohama, Japan}
\acmPrice{15.00}
\acmISBN{978-1-4503-XXXX-X/18/06}

\usepackage{caption}
\usepackage{subcaption}
\newcommand{\method}{{\textsc EvoK}} 
\begin{document}


% \title{Create a catchy title as the CHI'19 example paper}
\title{\method: Connecting loved ones through Heart Rate sharing}


\author{Esha Shandilya}
\authornotemark[1]
\email{es4524@rit.edu}
\affiliation{%
  \institution{Rochester Institute of Technology}
  \country{USA}
}

\author{Yiwen Wang}
\authornotemark[1]
\email{yw7615@rit.edu}
\affiliation{%
  \institution{Rochester Institute of Technology}
  \country{USA}
}

\author{Xuan Zhao}

\email{xz7320@rit.edu}
\affiliation{%
  \institution{Rochester Institute of Technology}
  \country{USA}
}

\authornote{equal contribution} %, names in $\alpha-\beta$

\author{Mingming Fan}
\email{mingming.fan@rit.edu}
\affiliation{%
  \institution{Rochester Institute of Technology}
  \country{USA}
}

\renewcommand{\shortauthors}{Shandilya and Wang, et al.}

\begin{abstract}
In this work, we present \method{}, a new way of sharing one's heart rate with feedback from their close contacts to alleviate social isolation and loneliness. \method{} consists of a pair of wearable prototype devices (i.e., sender and receiver). 
%\method allows users to share their heart rates with feedback from their close contacts to alleviate social isolation and loneliness. 
The sender is designed as a headband enabling continuous sensing of heart rate with aesthetic designs to maximize social acceptance.  The receiver is designed as a wristwatch enabling unobtrusive receiving of the loved one's continuous heart rate with multi-modal notification systems.
%and resuming the sender’s heart rate transmission.
% TODO: The Abstract should be within 150 words.
\end{abstract}





\maketitle

\section{Introduction}
% ACM's consolidated article template, introduced in 2017, provides a
% consistent \LaTeX\ style for use across ACM publications, and
% incorporates accessibility and metadata-extraction functionality
% necessary for future Digital Library endeavors. Numerous ACM and
% SIG-specific \LaTeX\ templates have been examined, and their unique
% features incorporated into this single new template.

% If you are new to publishing with ACM, this document is a valuable
% guide to the process of preparing your work for publication. If you
% have published with ACM before, this document provides insight and
% instruction into more recent changes to the article template.

% The ``\verb|acmart|'' document class can be used to prepare articles
% for any ACM publication --- conference or journal, and for any stage
% of publication, from review to final ``camera-ready'' copy, to the
% author's own version, with {\itshape very} few changes to the source.

%People’s mental health condition has always been a popular topic to study, increasing depression among the global population. 
According to WHO (World Health Organization), more than 264 million people worldwide have suffered from depression \cite{UND}.~Without any intervention, depression could lead to suicide and cause mortality in a hazardous condition. One factor that seriously affects people's mental health is social isolation and loneliness because of the lack of family ties, and communication \cite{amzat2016emotional}. Moreover, because of the COVID-19, the depression even doubled due to the implementation of new social rules such as social distancing and quarantine policy to protect everyone from the viruses \cite{marroquin2020mental}. The lack of interaction further intensifies their loneliness and adversely affects their mental health. In serious conditions, the isolated environment could induce physical illness and mental diseases~\cite{hawkley2010loneliness}. 

Researchers have investigated ways to alleviate the social loneliness of individuals, such as increasing one's interaction with their families and designing education programs and virtual companions~\cite{pettigrew2008addressing, tsiourti2014virtual, waycott2015ethics}. One way to improve one's social interaction and induce empathy is to sense and share their biosignals, such as breath patterns and heart rates \cite{frey2014heart, frey2018breeze, liu2019animo, min2014biosignal}. However, due to the limitation in comfort levels and elusive vibration feedback, it is necessary to investigate new wearable form factors to comfortably and effectively sense and share people's biosignals to increase their emotional connections with their loved ones. In this work, we design a pair of wearable devices for people to communicate their heart rate and for their loved ones to receive the heart rate notification with visual and audio feedback. The pair of devices is designed to strengthen the relationship between users and their loved ones through ambient feedback, comfortable and socially-acceptable design. 

\noindent {\bfseries Reproducibility:} The source code for \method{} is available at \url{https://github.com/EshaShandilya/evok}.

\section{Background and Related Work}
Biosignals, such as pulse sensing (PPG signal), brain electrical activity (EEG signal), and respiratory rate, are usually measured to understand the physiological process and activity of human beings \cite{frey2018breeze,ramirez2012detecting,temko2017accurate}. 
One such biosignal is heart rate, which is helpful in expressing distinct aspects of a person's emotional state \cite{slovak2012understanding}. Moreover, exchanging heart rate can affect users' social interaction and enhance their engagement \cite{frey2014heart,slovak2012understanding}.

This has motivated researchers to design different form factors to sense and share heart rates.
%Many types of design allows people to share their heart rate with others, aiming to increase connectedness across distance. 
For example, Werner et al. \cite{10.1145/1409240.1409338} designed a ring that can detect the wearer's heart rate and send it to the partner's ring via vibration feedback. However, participants felt the vibration feedback as a feeling of "electric shock" and was elusive to infer the corresponding heart rate. Croft and Lotan \cite{10.1145/1240866.1240936} created a device named imPulse with a curved surface so that users could put it on their laps and palms on the surface, providing synchronizing light and vibration feedback. However, imPulse is nonwearable, which makes it immobile and unusable when a user is performing other activities.
% and functional under the active status and not easy to carry around compares to wearable devices. 
Min and Nam designed WearBEAT to share body sound including the sounds of heartbeat \cite{min2014biosignal}. In their design, one user wore the sound input part on the chest mount to sense the heart rate, and the other user received vibration feedback on their wrist as output. However, the chest-mounted prototype violates the parameters of Zeagler's \cite{zeagler2017wear} body map locations for wearable; 
% according to Zeagler's \cite{zeagler2017wear} body map locations for wearable
since the position near breast could be uncomfortable for wearers especially for non-male users, which might affect its social-acceptability.
% affect the social acceptability and cause uncomfortable for female users on wearing. 

Inspired by these designs, in this work, we explore a different form factor to sense and share one's heart rate with the aim of increasing its comfort level and social-acceptability. We adopt PPG sensing to detect heart rate as it has been demonstrated to be a promising wearable heart rate sensing approach~\cite{frey2014heart}. 
%-- since heart rate is the most common biosignal  \cite{min2014biosignal} -- of an individual through a wearable technology to connect an isolated person with their friends and family. 
% We customize visual notification to transmit the normal and abnormal heart rate and continuously create a smoothing experience under the active and rest states. 
Through the combination of headband and wristwatch, we strive to provide a comfortable wearing experience and an intuitive interaction. 
% and heart rate is the most common biosignal  \cite{min2014biosignal}; we aim to detect and share the heart rate of an individual through a wearable technology to connect an isolated person with their friends and family. 
% Our prototype \method{} is designed to provide comfort, ambient notifications, and intuitive interactions while connecting the users.

% terms of the comfort level, providing ambient notifications, and socially-acceptable while connecting two users. 

% This is an example citation~\cite{Abril07}. Remember to add the citation's bibtex into the bibs/main.bib file.
\section{\protect \method}

\subsection{Design Considerations}
% updated
We followed an iterative design process to finalize the key design considerations for our wearable devices. The design considerations were based on - i) trade-offs between the comfort level and the detection accuracy of the heart rate sensor, ii) ambient notification, iii) socially-acceptable designs. %, which helped us evolve and inform the final design of the wearable devices.
\\
\textbf{Trade-offs between the Comfort Level and the Detection Accuracy of the Heart Rate Sensor:} Before concluding the sensor’s placement, we tested various feasible locations, specifically the finger-tip and the ear-lobe, to wear the heart rate sensor to get accurate heart rate. We observed that the finger-tip position added noise to the heart rate due to hand-movements, whereas the sensor’s placement on the ear-lobe reduced noise in the recorded heart rate. Consequently, we decided to use the ear-lobe for the sensor’s placement. 
However, the batteries’ weight pulls the sensor down when placed on the earlobe, impacting the wearer’s comfort and the sensor’s accuracy. To overcome this challenge, we brainstormed a solution that could support the batteries’ weight to keep the heart rate sensor’s position intact on the ear lobe, providing a seamless experience to the wearer. We deliberated the feasibility of multiple design alternatives such as \textit{earrings}, \textit{neckwear}, \textit{hair clips}, and \textit{headbands}. The headband prototype, Figures \ref{fig:fig2}, offers the best weight distribution of the batteries than the competing alternatives; the encased batteries are attached to the headband, which rests on the head, providing stable heart rate sensor positioning.\\
\textbf{Ambient Notification:} The next design consideration was to design an unobtrusive notification, which the wearer can easily follow without getting overwhelmed with the constant influx of the sender’s heart rate. According to Hansson and Ljungstrand \cite{hansson2000reminder} colored LED lights are less intrusive methods of notification systems than other forms such as sound. Therefore, we devised three different LED lights to indicate the heart rate range; considering that the normal heart rate range is between 60 to 100 for an adult, according to Mayo Clinic \cite{Mayo}. In the Figures \ref{fig:fig7}, we see the blue light for heart rate less than 60, the green light for the heart rate between 60 and 100, and red to highlight the heart rate beyond 100. In case there is a constant high heart rate of the sender (beyond 100), the receiver (wearer) will be alerted by a high pitched sound. Moreover, we also provide an option for the user to control the sender’s heart rate transmission by pressing a button on the receiver’s prototype. We abandoned the idea of incorporating haptic feedback to the prototypes, as it may be intrusive and dysfunctional for users to interact with the prototype.\\
\textbf{Socially-acceptable Designs:} Social acceptability of a wearable device directly depends on its placement on the user's body~\cite{zeagler2017wear}. Our wearable designs are conceptualized according to the socially-acceptable body locations suggested by Zeagler \cite{zeagler2017wear} that provide comfort and confidence to the wearer in public.
Therefore, after referring to the body map \cite{zeagler2017wear}, we chose the two areas -- the head and the wrist that offer comparatively better affordance for the device placement, and thus we select the form-factors of a headband for the sender and a wrist-watch for a receiver.
Such form-factors,  (headband and wrist-watch), Figures \ref{fig:fig2} and \ref{fig:fig6}, are intuitive, user-friendly and easy to interact with as these are familiar form-factors to the users. We carefully determined these designs as these are gender-neutral and to the most extent used by many. However, we did not conduct user research to assess the designs' social acceptance.

\begin{figure}[h]
  \centering
  \includegraphics[width=\linewidth]{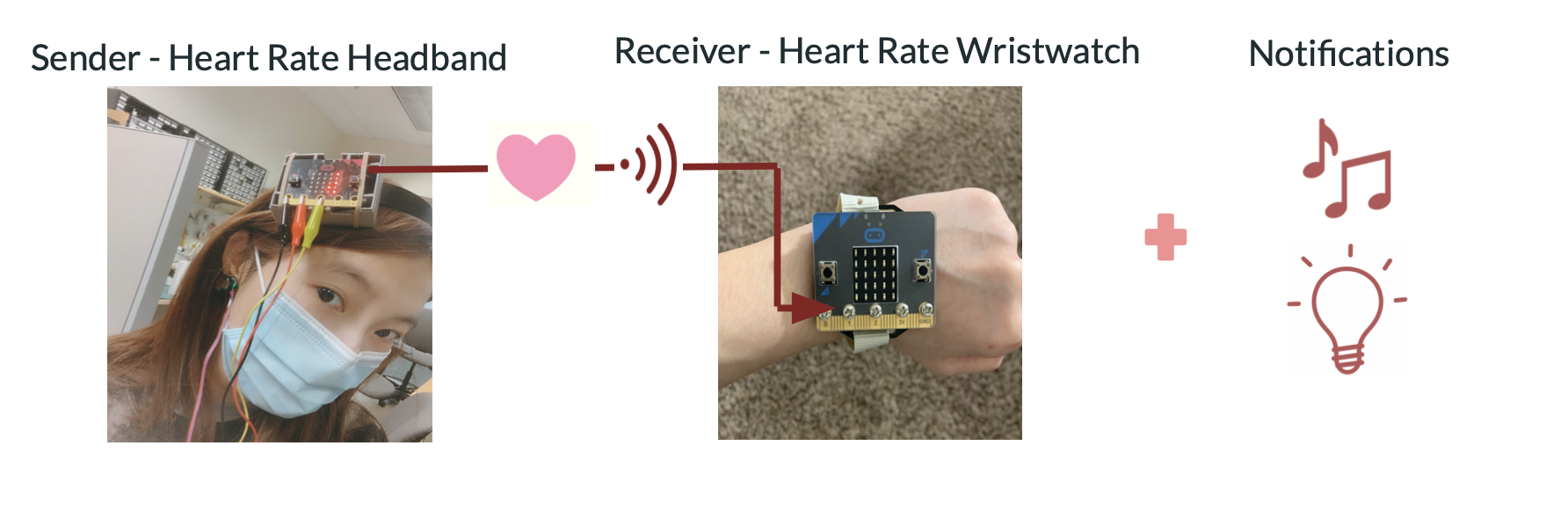}
  \caption{This is the final stage, where we work to consolidate all the sensors in a usable and functional wearable design. Here the heart rate sensor is a headband design for the sender and when heart rate will be sensed from a specific range then, the data will be shared in the form of visuals with music to the receiver which is in the form of a wrist watch. For different heart ranges, there are specific notifications, please refer Figure \ref{fig:fig7}  for full explanation on design notifications.}
  \Description{The headband and other components.}
  \label{fig:fig3}
\end{figure}

\subsection{Implementation}
Our system consists of two parts, the sender and the receiver. The sender's heart rate will be detected and sent to the receiver, and the receiver will get different visual and audio notifications according to the received heart rate value.  

\begin{figure*}
     \centering
     \begin{subfigure}[b]{0.28\textwidth}
         \centering
         \includegraphics[width=\textwidth]{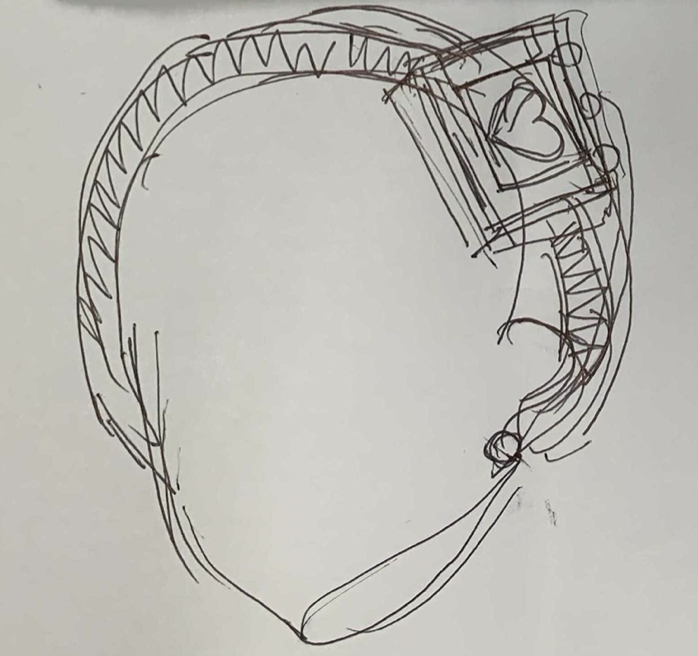}
     \end{subfigure}
     \hfill
     \begin{subfigure}[b]{0.57\textwidth}
         \centering
         \includegraphics[width=\textwidth]{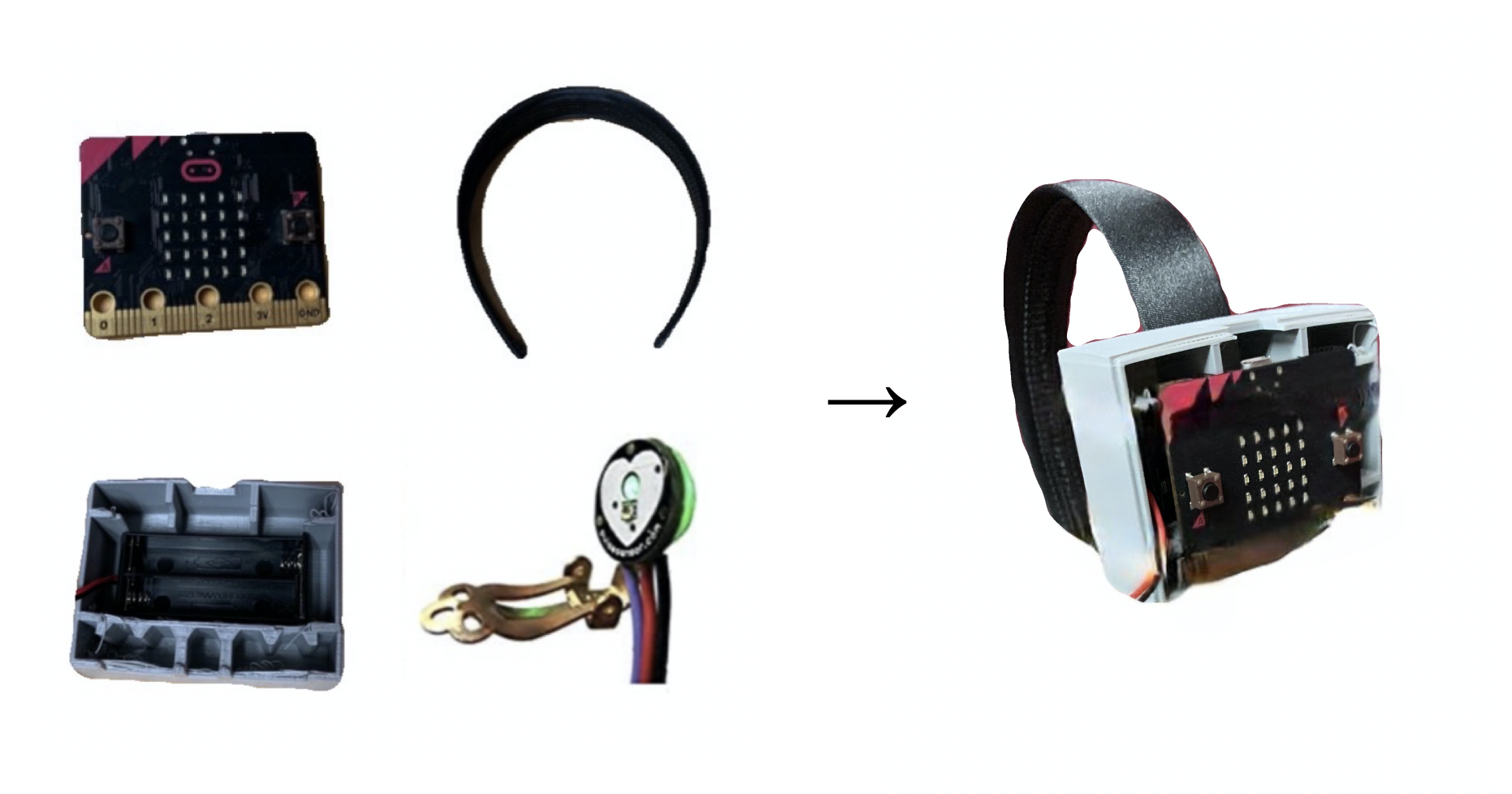}
     \end{subfigure}
      \caption{The final design for the sender - Sender's headband: a wearable headband banded with a 3D printing compact box, includes battery, Micobits, connector and linked with pulse sensor.}
  \Description{The headband and other components.}
  \label{fig:fig2}
\end{figure*}

To conceptualize the \textbf{sender} part, we used a pulse rate sensor to detect the heart rate. The pulse rate sensor was connected to the micro: bit and the code for calculating the heart rate was downloaded to the micro: bit. The pulse rate sensor could be placed on the fingertip or earlobe. We tested both these two placements and found that placing the sensor on the earlobe gave us more stable signals. Also, considering the convenience of a user wearing the wearable device, the sensor should be intact while working out, performing some activity, and resting state. Thus, we decided to put the sensor on the earlobe. We tried to put the sensor on the earlobe and connected it to the micro: bit. We found that the gravity of the micro: bit would exert a great downward force on the earlobe, which could cause ear discomfort. We needed to put the micro: bit in a supportive position. Our initial idea was to put it in the cloth pocket; however, not all tops had pockets. Then we turned to body parts and found the head would be a good choice to put the micro: bit on. We wanted to make our device portable and could be used by anyone. The idea of the hairpin was abandoned because it did not apply to people with short hair. Our idea was to attach the micro: bit to a headband Figure~\ref{fig:fig2}. To compact the battery and connector, we used the 3D printer to print a small box so that all components can get packed in one Figure~\ref{fig:fig2}. When the sender uses this device, they need to put the headband on and place the pulse rate sensor on the earlobe, as shown in the Figure~\ref{fig:fig3}. After wearing the device, it needs to collect one or two minutes of data before getting the wearer's normal heart rate.

% \begin{figure}[h]
%   \centering
%   \includegraphics[width=0.5\linewidth]{figures/figure2.jpg}
%   \caption{The final design for the sender - Sender's headband: a wearable headband banded with a 3D printing compact box, includes battery, Micobits, connector and linked with pulse sensor.}
%   \Description{The headband and other components.}
%   \label{fig:fig2}
% \end{figure}

% \begin{figure}[h]
%   \centering
%   \includegraphics[width=0.69\linewidth]{figures/figure9.png}
%   \caption{The final design for the sender - Sender's headband: a wearable headband banded with a 3D printing compact box, includes battery, Micobits, connector and linked with pulse sensor.}
%   \Description{The headband and other components.}
%   \label{fig:fig9}
% \end{figure}

\begin{figure}[h]
  \centering
  \includegraphics[width=0.6\linewidth]{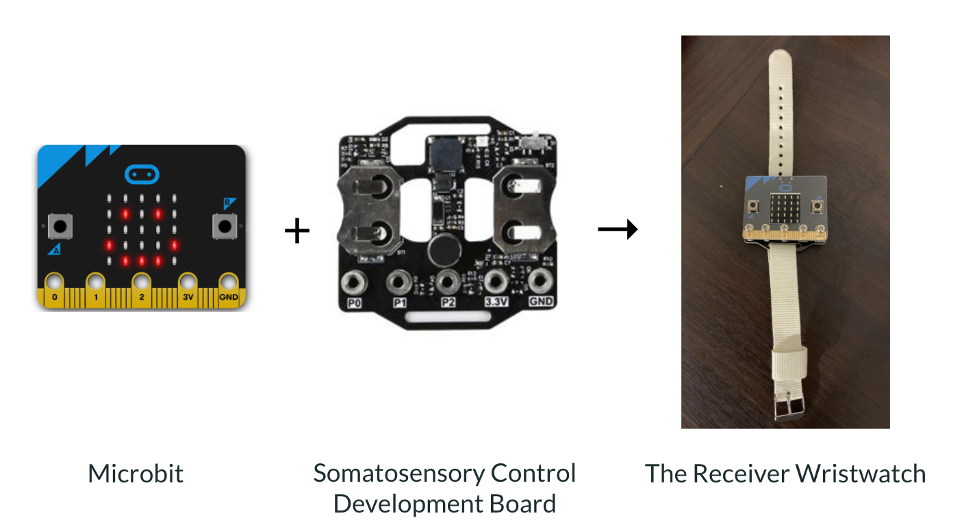}
  \caption{Receiver's wristwatch: contains a somatosensory control development board which enables a user to interact with the prototype through audio, visual and haptic feedback.}
  \Description{Receiver's wristwatch components.}
  \label{fig:fig6}
\end{figure}

\begin{figure*}
     \centering
     \begin{subfigure}[b]{0.5\textwidth}
         \centering
         \includegraphics[width=\textwidth]{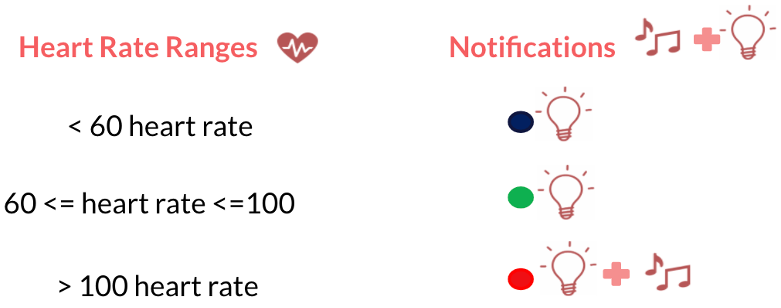}
     \end{subfigure}
     \hfill
     \begin{subfigure}[b]{0.45\textwidth}
         \centering
         \includegraphics[width=\textwidth]{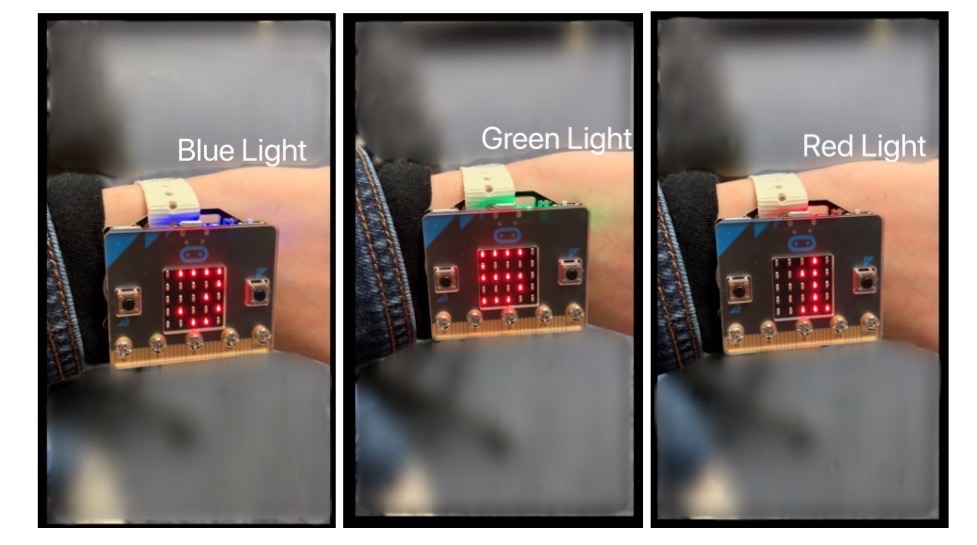}
     \end{subfigure}
  \caption{Three ranges of heart rate with corresponding feedback: Blue LED represents less than 60. Green LED represents the normal range between 60 and 100. Red LED and alarming sound represents over 100. According to Mayo Clinic \cite{Mayo}
  , the normal resting heart rate of an individual ranges between 60 and 100; other factors such as, age, fitness level, emotions could also influence the heart rate.
  The heartbeat value of the sender is displayed on the Microbit's screen. Note: The heart rates displayed are two and three digit numbers. One single digit is shown at a time, and the digits are moving from right to left. The first value on the receiver's device is the first digit 3 of 30 with a blue LED light, second value is 5, the last digit of value 65 with green LED light, and the last value is 1, the first digit of 130 with red LED light.}
  \Description{Different LED light colors and audio for different heart beat value.}
  \label{fig:fig7}
\end{figure*}

For the \textbf{receiver} part, we designed some feedback and interactions for the receiver. The receiver also needed to wear the micro: bit to receive the data from the sender. To make it easier to see the heart rate value and feedback, we made the device in the form of a wristwatch. We purchased a somatosensory control development board with a wrist band, an RGB led light, a buzzer, and a speaker, Figure \ref{fig:fig6}. The micro: bit was connected to the board Figure \ref{fig:fig6}. We tried to utilize the buzzer to provide haptic feedback; however, the micro: bit was not powerful enough to implement real-time haptic feedback and caused long delays. In our design, we only used the led light and speaker to provide visual and audible feedback, the description can be seen in Figure \ref{fig:fig7}. When the sender’s heart rate was below the normal range \cite{Mayo}, which was less than 60 per minute, the light would be blue. When the heart rate value was in the normal range between 60 and 100, the light would be green. When the heart rate was beyond the normal range, which meant over 100 per minute, the light would turn red, accompanied by a beep sound. The normal range of 60 to 100 was only used for our test. Users could set their own heart rate normal range. Additionally, to provide more flexibility, the receiver could control whether to receive the data or not. They could press the left button on the micro: bit to stop receiving the data and feedback and press the left button again to resume receiving the data and feedback. 

\section{Conclusion}
We present novel prototype designs with feedback for sharing heart rates to alleviate people’s social isolation from their loved ones. This work presents a demonstration with a pair of wearable devices that mainly use micro:bit processor and a heart rate sensor. We exclusively made our designs wearable to ensure continuous connectivity through heart rate sharing and indicating the user’s physical and mental well-being. 

Future work should investigate users’ attitudes and perception towards the proposed way of sensing and sharing heart rates and elicit feedback to further improve its comfort level and social acceptability. Moreover, it is worth conducting a long-term in-the-wild study to reveal both technical issues, such as battery life, and practical issues emerging from a wide range of daily scenarios. Lastly, as heart rate patterns may vary from person to person and can carry important health- or emotion- related information, it is worth exploring the characteristics of such patterns and designing prototypes to capture and communicate them among loved ones.

\bibliographystyle{ACM-Reference-Format}
\bibliography{bibs/main,bibs/references}

% %%
% %% If your work has an appendix, this is the place to put it.
% \appendix









\end{document}